\documentclass[12pt]{elsart}
\usepackage{amsmath,amssymb,euscript,bbold}
\usepackage{array}
\usepackage{fancybox}
\catcode`\@=11
\def\@journal{Nuclear Physics B}
\catcode`\@=12
\newcommand{\EE}{\mathbb{E}}
\newcommand{\MM}{\mathbb{M}}

\newcommand{\RR}{\mathbb{R}}
\newcommand{\fg}{\mathfrak{g}}
\newcommand{\Spin}{\mathrm{Spin}}
\newcommand{\SO}{\mathrm{SO}}
\newcommand{\SU}{\mathrm{SU}}
\newcommand{\SL}{\mathrm{SL}}
\newcommand{\sC}{\mathsf{C}}
\newcommand{\sP}{\mathsf{P}}
\newcommand{\eL}{\EuScript{L}}
\newcommand{\eV}{\EuScript{V}}
\newcommand{\1}{\mathbb{1}}
\newcommand{\Cl}{\mathrm{C}\ell}
\newcommand{\Mat}{\mathrm{Mat}}
\newcommand{\Tr}{\mathrm{Tr}}
\newcommand{\Det}{\mathrm{Det}}
\newcommand{\dd}{\mathbb{d}}
\newcommand{\CS}{\mathsf{CS}}
\newcommand{\II}{\ensuremath{\mathrm {I\!I}}}
\renewcommand{\Sp}{\mathrm{Sp}}
\renewcommand{\d}{\partial}
\begin{document}

\begin{frontmatter}
\title{Euclidean D-branes and higher-dimensional gauge theory}
\author{BS~Acharya\thanksref{emailbobby}\thanksref{PPARCb}},
\author{JM~Figueroa-O'Farrill\thanksref{emailjmf}\thanksref{EPSRCj}},
\author{B~Spence\thanksref{emailbill}\thanksref{EPSRCb}}
\address{Department of Physics, Queen Mary and Westfield College,
Mile End Road, London E1 4NS, UK}
and
\author{M~O'Loughlin\thanksref{emailmjol}}
\address{International Centre for Theoretical Physics, 34100,
Trieste, Italy}
\thanks[emailbobby]{\tt mailto:r.acharya@qmw.ac.uk}
\thanks[PPARCb]{Supported by PPARC.}
\thanks[emailjmf]{\tt mailto:j.m.figueroa@qmw.ac.uk}
\thanks[EPSRCj]{Supported by the EPSRC under contract GR/K57824.}
\thanks[emailbill]{\tt mailto:b.spence@qmw.ac.uk}
\thanks[EPSRCb]{EPSRC Advanced Fellow.}
\thanks[emailmjol]{\tt mailto:mjol@ictp.trieste.it}
\begin{abstract}
We consider euclidean D-branes wrapping around manifolds of
exceptional holonomy in dimensions seven and eight.  The resulting
theory on the D-brane---that is, the dimensional reduction of
10-dimensional supersymmetric Yang--Mills theory---is a cohomological
field theory which describes the topology of the moduli space of
instantons.  The 7-dimensional theory is an $N_T{=}2$ (or balanced)
cohomological theory given by an action potential of Chern--Simons
type.  As a by-product of this method, we construct a related
cohomological field theory which describes the monopole moduli space
on a 7-manifold of $G_2$ holonomy.  
\end{abstract}
\end{frontmatter}

\section{Introduction}

D-branes (see for instance \cite{PolTASI}) lead naturally to the study
of geometries in dimension greater than four, and in particular to the
calibrated geometries introduced in \cite{HL}.  Part of the BPS
spectrum of M-theory or superstring theory compactified down to
realistic dimensions consists of branes wrapping around supersymmetric
cycles \cite{BBS,OOY,BBMOOY}, which are precisely the calibrated
submanifolds \cite{BSV}.  Calibrated geometries are particularly rich
in dimensions six, seven and eight, and precisely on riemannian
manifolds admitting parallel spinors \cite{LM,H}; in other words, on
manifolds of reduced holonomy.

These very manifolds also play a crucial role in the ``Oxford
programme'' \cite{DT} to generalise Donaldson--Floer--Witten theory to
higher dimensions.  The Yang--Mills equations on these manifolds admit
instanton-like solutions, obtained by imposing linear constraints on
the Yang--Mills curvature \cite{CDFN,Ward} which simply project it
onto a particular irreducible representation of the holonomy group.
Like in four dimensions, the Yang--Mills action on these manifolds
satisfies an $L^2$ bound which the instantons saturate \cite{AOL}.
In contrast with the familiar 4-dimensional case, very little is
known about the higher dimensional instantons and in particular about
their moduli space. One particularly useful approach to the study of
4-dimensional instantons is via topological field theory
\cite{WittenTFT} and it is hoped that a similar approach might prove
fruitful also in higher dimensions.

Indeed higher-dimensional generalisations of Witten's original
cohomological field theory were written down independently in
\cite{BKS} and \cite{AOS}.  Unlike the 4-dimensional case, these
cohomological field theories are not topological, since they depend
upon the reduction of the holonomy group; but as shown in \cite{AOS}
the observables are locally constant in the space of metrics of
reduced holonomy.\footnote{It is not known whether this space is
connected or, if not, whether these invariants detect the connected
component. We are grateful to IM~Singer for some correspondence on
this point.} It was remarked in both of these papers that these
theories had precisely the same spectrum as 10-dimensional
supersymmetric Yang--Mills reduced to the corresponding dimension.
Thus it was conjectured that these theories should be obtained in this
way.  The point of this paper is to prove this conjecture.

The dimensional reduction of 10-dimensional supersymmetric Yang--Mills
is of course an old game.  This is used, for example, to explain the
existence of extended supersymmetric Yang--Mills theories in four
dimensions and gives the simplest known method to construct the
actions.  Much less studied are the reductions to euclidean spaces or
more generally to riemannian manifolds.  Results in this direction
have been obtained in \cite{BSV,BTNT2too}, who considered euclidean
D-branes wrapping around calibrated submanifolds.  The resulting
theories on the D-brane were seen to be topologically twisted
Yang--Mills theory, with the components of the 10-dimensional gauge
field in directions normal to the D-brane being no longer simply
scalars, but rather sections of the normal bundle to the calibrated
submanifold.

The theories we will describe in what follows can be understood as
those arising out of euclidean D-branes wrapping around the full
manifold, it being trivially calibrated by the volume form.  Our
approach is the following.  We start with 10-dimensional
supersymmetric Yang--Mills theory and reduce it to $d$-dimensional
euclidean space, where for the purposes of this paper $d{=}7,8$.  The
resulting lagrangian can be promoted to any spin manifold $M$ of
dimension $d$ by simply covariantising the derivatives with respect to
the spin connection; but the supersymmetry transformations will fail
to be a symmetry unless the parameters are covariantly constant.  This
requires that $M$ admit parallel spinors, and this singles out those
manifolds admitting a reduction of the holonomy group to a subgroup of
$\Spin(7)$ in eight dimensions and to a subgroup of $G_2$ in seven.
Covariance of the supersymmetry algebra under the holonomy group
implies that the commutator of two supersymmetry transformations with
parallel spinors as parameters will result (at least on shell and up
to gauge transformations) in a translation by a parallel vector.
Since for the irreducible manifolds we consider there are no such
vectors, the supersymmetry transformation is a BRST symmetry.
Therefore the resulting theory is cohomological.  As we will see, the
theories constructed in this paper in this fashion agree morally with
the ones considered in \cite{BKS} and \cite{AOS}, whose observables
are topological invariants of the moduli space of instantons.

As a by-product of our construction we also arrive at a related
cohomological field theory, briefly discussed in \cite{BKS}, which
describes what could be termed the monopole moduli space in a
7-manifold of $G_2$ holonomy.  Just like monopoles in three
dimensions---by which we mean any solution of the Bogomol'nyi
equation---can be understood as 4-dimensional instantons with a
certain symmetry, every 7-dimensional monopole yields a particular
solution of the 8-dimensional instanton equations.  Their
cohomological field theory is therefore obtained from the one
describing the 8-dimensional instantons.  In this way, we can
understand this theory as a natural higher-dimensional generalisation
of the cohomological field theory written down in \cite{BRT,BG}.

The plan of this paper is the following.  After a cursory look at our
notation, we briefly review 10-dimensional supersymmetric Yang--Mills
theory in Section 2.  In Section 3 we discuss the reduction to
8-dimensional euclidean space, and then to manifolds of $\Spin(7)$
holonomy.  In Section 4 we consider the reductions to
7-dimensional euclidean space and then to manifolds of $G_2$
holonomy.  Finally in Section 5 we will offer some conclusions.

After completion of this work there appeared the paper \cite{BTESYM}
which also deals with the present topic, but (thankfully!) in a
largely complementary manner.

\subsection{A word on notation}

Throughout this paper we will use the notation $\MM^{s{+}t}$ to refer
to ($s{+}t$)-di\-men\-sion\-al Minkowski spacetime.  In addition
$\EE^d = \MM^{d{+}0}$ will denote $d$-dimensional euclidean space.
Spinor notation follows the conventions in \cite{LM}.  In particular,
$\Cl(s,t)$ denotes the Clifford algebra (notice the sign!)
\begin{equation*}
\Gamma_\mu \Gamma_\nu + \Gamma_\nu \Gamma_\mu = -2 \eta_{\mu\nu} \1~,
\end{equation*}
where $\eta_{\mu\nu}$ is diagonal with signature $+s{-}t$.  Our
notation for representations of the spin groups is the following.
The trivial, vector and adjoint representations are denoted
$\bigwedge^0$, $\bigwedge^1$, and $\bigwedge^2$ respectively.  For
example, for $\Spin(7)$ these are the $\mathbf{1}$, $\mathbf{7}$ and
$\mathbf{21}$; and for $\Spin(8)$ the $\mathbf{1}$, $\mathbf{8_v}$ and
$\mathbf{28}$.  The half-spin representations are denoted $\Delta$ for
odd-dimensional spin groups and $\Delta_\pm$ for the even-dimensional
spin groups.  For example, for $\Spin(7)$, $\Delta$ is the
$\mathbf{8}$, whereas for $\Spin(8)$, $\Delta_+$ and $\Delta_-$ are
the $\mathbf{8_s}$ and $\mathbf{8_c}$ respectively.  Other group
theory notation will be introduced as needed.

\section{Ten dimensions}

We start by briefly reviewing supersymmetric Yang--Mills in
10-dimensional Minkowski spacetime $\MM^{9{+}1}$.

Let $\Hat\Gamma_\mu$ denote the generators of the Clifford algebra
$\Cl(9,1)$.  As real associative algebras $\Cl(9,1) \cong
\Mat_{32}(\RR)$, whence the $\Hat\Gamma_\mu$ can be chosen to be
$32\times32$ real matrices.  This means that $\Hat\Gamma_0$ can be
chosen in addition to be symmetric and $\Hat\Gamma_i$ for
$i=1,\ldots,9$ antisymmetric.  The charge conjugation matrix $\Hat\sC$
satisfies $\Hat\sC^t = - \Hat\sC$, and also $\Hat\Gamma_\mu^t = -
\Hat\sC \Hat\Gamma_\mu\Hat\sC^{-1}$.  This means that it anticommutes
with $\Hat\Gamma_0$ and commutes with all the other $\Hat\Gamma_i$.
Therefore we can choose it to be
\begin{equation}\label{eq:9+1C}
\Hat\sC = \Hat\Gamma_1 \cdots \Hat\Gamma_9~.
\end{equation}
One can check that indeed $\Hat\sC^t = - \Hat\sC$.  The chirality
operator $\Hat\Gamma_{11}$ is defined by
$\Hat\Gamma_0\Hat\Gamma_1\cdots \Hat\Gamma_9$.  One can check that
$\Hat\Gamma_{11}^2 = \1$, whence in this realisation it is both real
and symmetric.  In other words, Majorana--Weyl spinors exist in
$\MM^{9{+}1}$.

The action for supersymmetric Yang--Mills theory in $9{+}1$
dimensions can be formulated in terms of a gauge field $A_\mu$ and a
negative chirality Majorana--Weyl spinor $\Psi$, taking values in a
Lie algebra $\fg$ assumed to possess an invariant metric.  The
lagrangian density is given by
\begin{equation}\label{eq:9+1SYM}
\eL = \tfrac14 (F_{\mu\nu}, F^{\mu\nu}) + \tfrac{i}{2} (\Bar\Psi,
\Hat\Gamma^\mu D_\mu \Psi)~,
\end{equation}
where
\begin{itemize}
\item $F_{\mu\nu} = \d_\mu A_\nu - \d_\nu A_\mu -  [A_\mu,A_\nu]$;
\item $D_\mu \Psi = \d_\mu\Psi -  [A_\mu,\Psi]$; and
\item $(-,-)$ is a fixed invariant metric on the Lie algebra $\fg$.
\end{itemize}
This action is invariant under gauge transformations and under the
super Poincar\'e group in $9{+}1$ dimensions.  Infinitesimally, the
gauge transformations take the form:
\begin{equation}\label{eq:9+1gauge}
\delta_\omega A_\mu = D_\mu\omega \qquad \text{and}\qquad
\delta_\omega \Psi = [\omega,\Psi]~,
\end{equation}
where $\omega$ is a Lie-algebra valued function.  The supersymmetry
transformations are given by
\begin{equation}\label{eq:9+1SUSY}
\delta_\varepsilon A_\mu = i\Bar\varepsilon \Hat\Gamma_\mu \Psi
\qquad\text{and}\qquad
\delta_\varepsilon \Psi = \tfrac12 F^{\mu\nu} \Hat\Gamma_{\mu\nu}
\varepsilon~,
\end{equation}
where $\varepsilon$ is a constant negative-chirality Majorana--Weyl
spinor and $\Hat\Gamma_{\mu\nu} = \tfrac12
(\Hat\Gamma_\mu\Hat\Gamma_\nu - \Hat\Gamma_\nu\Hat\Gamma_\mu)$.
There are eight bosonic and eight fermionic real physical degrees of
freedom, but because the bosonic and fermionic off-shell degrees of
freedom do not match in this formulation, the supersymmetry algebra
will only close on shell and, indeed, up to gauge transformations.

\section{Eight dimensions}

In this section we derive a supersymmetric Yang--Mills theory in
$\EE^8$ by dimensional reduction from $\MM^{9{+}1}$ and then we will
extend it to a riemannian manifold $M$ of $\Spin(7)$ holonomy.  The
resulting action defines a cohomological theory whose observables are
topological invariants of the moduli space of instantons on $M$.

\subsection{Properties of spinors}\label{sec:8spinors}

In order to facilitate the dimensional reduction we will first choose
a convenient realisation of the $\Gamma$-matrices in ten dimensions.
The Clifford algebra isomorphism $\Cl(9,1) \cong \Cl(8,0) \otimes
\Cl(1,1)$ suggests one such realisation.  We let the Clifford algebra
$\Cl(1,1)$ be generated by $\sigma_1$ and $-i\sigma_2$.  The chirality
operator is given by $\sigma_3$.  Notice that these matrices are real
and the chirality operator is diagonal, indicative of the existence of
Majorana--Weyl spinors in $1{+}1$ dimensions.  Let $\Gamma_i$ for
$i=1,\ldots,8$ denote the $\Gamma$-matrices in eight dimensions.
They are $16\times 16$ matrices which can be chosen to be real and
antisymmetric.  Moreover since Majorana--Weyl spinors also exist in
$8{+}0$ dimensions, we can take $\Gamma_9 \equiv \Gamma_1 \cdots
\Gamma_8$ to be diagonal.  Now consider the following expressions:
\begin{equation}\label{eq:explicitGammas}
\Hat\Gamma_0 = \1 \otimes \sigma_1 \qquad
\Hat\Gamma_i = \Gamma_i \otimes \sigma_3\qquad
\Hat\Gamma_9 = \1 \otimes (-i\sigma_2)~.
\end{equation}
These are $\Gamma$-matrices for $\Cl(9,1)$.  Notice that they are real
and antisymmetric, except for $\Hat\Gamma_0$ which is real and
symmetric.  In this realisation, the charge conjugation matrix
$\Hat\sC$ given by \eqref{eq:9+1C} becomes $\Hat\sC = \Gamma_9 \otimes
(-i\sigma_2)$.  Notice that $\Gamma_9$ can be identified with the
charge conjugation matrix $\sC$ in $8{+}0$ dimensions, since it is
symmetric and anticommutes with the $\Gamma_i$.  In this realisation,
the chirality operator in $9{+}1$ dimensions is given by
$\Hat\Gamma_{11} = \Gamma_9 \otimes \sigma_3$.

Now let $\Psi$ be a 32-component spinor in $9{+}1$ dimensions.  In
terms of the above decomposition of $\Cl(9,1)$ we can write it as
follows: $\Psi = (\psi_1~ \psi_2)^t$ where $\psi_A$ are 16-component
spinors acted on by the $\Gamma_i$.  Suppose that $\Psi$ is chiral:
$\Hat\Gamma_{11} \Psi = \pm \Psi$.  Then the components $\psi_A$ are
chiral with respect with $\Gamma_9$.  Indeed $\Gamma_9 \psi_1 = \pm
\psi_1$ and $\Gamma_9 \psi_2 = \mp \psi_2$.  On the other hand, if
$\Psi$ is Majorana: $\Bar\Psi \equiv \Psi^\dagger \Hat\Gamma_0 =
\Psi^t \Hat\sC$, then $\psi_A$ satisfy the following  reality
conditions: $\psi_1^* = -\Gamma_9 \psi_1$ and $\psi_2^* = \Gamma_9
\psi_2$.  Therefore, if $\Psi$ is Majorana--Weyl then $\psi_A$ are
either both real or both imaginary according to whether $\Psi$ has
negative or positive chirality, respectively.  In our case, we have
chosen the spinor $\Psi$ in \eqref{eq:9+1SYM} to have negative
chirality, whence $\psi_A$ are real.

\subsection{Dimensional Reduction}

We now dimensionally reduce to $\EE^8$ by simply dropping the
dependence on $x^0$ and $x^9$.  In other words, we let $\d_0 = \d_9 =
0$.  The Lorentz symmetry of the theory is therefore broken down to
$\SO(8)\times\SO(1,1)$, which suggests the following decomposition of
the fields: $A_\mu = (A_i,A_9,A_0)$ and $\Psi = (\psi_1~\psi_2)^t$,
where $\psi_1$ and $\psi_2$ are respectively negative and positive
chirality spinors of $\Spin(8)$, and $A_9$ and $A_0$ are scalars.  Of
course, all fields remain Lie algebra valued.

We first tackle the bosonic part of the action.  It is enough to
realise that $F_{0i} = -D_i A_0$, $F_{9i} = -D_i  A_9$ and $F_{09} = [
A_9, A_0]$, to to derive
\begin{equation*}
\tfrac14 (F_{\mu\nu},F^{\mu\nu}) = \tfrac14 \|F_{ij}\|^2 + \half
\|D_i A_9\|^2 - \half \|D_i A_0\|^2 - \half
\|[ A_9, A_0]\|^2~.
\end{equation*}
Similarly, notice that $D_0\Psi = - [ A_0,\Psi]$ and $D_9\Psi = - [
A_9,\Psi]$.  Using this and the explicit realisation of the
$\Hat\Gamma$-matrices given by \eqref{eq:explicitGammas} one obtains
(with $\epsilon^{12}= - \epsilon^{21} = 1$)
\begin{multline*}
\tfrac{i}{2} (\Bar\Psi,\Hat\Gamma^\mu D_\mu \Psi) = 
- \tfrac{i}{2} \epsilon^{AB} (\psi_A^t,\Gamma_i D_i \psi_B)\\
- \tfrac{i}{2} (\psi_1^t, [ A_9- A_0,\psi_1])
+ \tfrac{i}{2} (\psi_2^t, [ A_9+ A_0,\psi_2])~.
\end{multline*}
This suggests that we define fields $\phi_\pm \equiv  A_9\pm A_0$.  In
terms of these fields the dimensionally reduced lagrangian density
becomes:
\begin{multline}\label{eq:action8}
\eL = \tfrac14 \|F_{ij}\|^2 + \half (D_i\phi_+,D_i\phi_-) - \tfrac18
\|[\phi_+,\phi_-]\|^2\\ - \tfrac{i}{2} \epsilon^{AB}
(\psi_A^t,\Gamma_i D_i \psi_B) - \tfrac{i}{2} (\psi_1^t,
[\phi_-,\psi_1]) + \tfrac{i}{2} (\psi_2^t, [\phi_+,\psi_2])~.
\end{multline}
The R-symmetry $\SO(1,1)$ acts diagonally on the fields
$(A_i,\phi_+,\phi_-,\psi_1,\psi_2)$ with weights
$(0,+1,-1,+\half,-\half)$.  Finally, using:
\begin{equation*}
\Hat\Gamma_{ij} = \Gamma_{ij}\otimes\1\quad
\Hat\Gamma_{0i} = \Gamma_i\otimes(-i\sigma_2)\quad
\Hat\Gamma_{9i} = \Gamma_i\otimes\sigma_1\quad
\Hat\Gamma_{09} = \1\otimes\sigma_3~,
\end{equation*}
the supersymmetry transformations take the form:
\begin{align}\label{eq:susytrans}
\delta_\varepsilon A_i &= -i \epsilon^{AB} \varepsilon_A^t \Gamma_i
\psi_B\notag\\
\delta_\varepsilon \phi_+ & = 2 i \varepsilon^t_1 \psi_1\notag\\
\delta_\varepsilon \phi_- & = -2 i \varepsilon^t_2
\psi_2\notag\\
\delta_\varepsilon \psi_1 & = \half \left( F_{ij}\Gamma_{ij} +
[\phi_+,\phi_-]\right) \varepsilon_1 - D_i\phi_+ \Gamma_i
\varepsilon_2\notag\\
\delta_\varepsilon \psi_2 & = \half \left( F_{ij}\Gamma_{ij} - 
[\phi_+,\phi_-]\right) \varepsilon_2 - D_i\phi_- \Gamma_i
\varepsilon_1~.
\end{align}

\subsection{Manifolds of exceptional holonomy}

A natural question to ask is whether this theory can be defined on
8-di\-men\-sion\-al manifolds other than $\EE^8$.  The action defined
by \eqref{eq:action8} makes sense on an arbitrary spin manifold
provided that we redefine the covariant derivative on the fermions to
include the spin connection.  However it will {\em not\/} be invariant
under the supersymmetry transformations unless the spinor parameters
$\varepsilon_A$ are covariantly constant with respect to the spin
connection.  We are therefore led to ask on which 8-dimensional spin
manifolds do parallel spinors exist.  Of necessity such manifolds must
have reduced holonomy $H \subset \SO(8)$, since the spinor
representations of $\Spin(8)$ must contain a singlet when decomposed
under $H$.  The subgroups of $\Spin(8)$ which leave invariant a spinor
are all subgroups of a $\Spin(7)$ subgroup.  Those under which the
manifold remains irreducible are $\Spin(7) \supset \SU(4) \supset
\Sp(2)$.  The latter two groups correspond to Calabi--Yau 4-folds and
hyperk\"ahler manifolds respectively.  Manifolds of $\Spin(7)$
holonomy possess one parallel spinor, whose chirality depends on which
$\Spin(7)$ subgroup of $\SO(8)$ we choose.  There are three
inequivalent $\Spin(7)$ subgroups of $\SO(8)$ related by triality.
One of these conjugacy classes does not lead to parallel spinors, but
the other two do.  Calabi--Yau 4-folds possess two parallel spinors of
the same chirality, determined by the conjugacy class of the $\SU(4)$
subgroup of $\SO(8)$ or equivalently by the conjugacy class of the
$\Spin(7)$ subgroup to which the $\SU(4)$ belongs.  A similar story
holds for hyperk\"ahler manifolds, which possess three parallel
spinors of the same chirality.

Although similar results to those we are about to describe hold on
Calabi--Yau 4-folds and hyperk\"ahler manifolds, we will focus in this
paper on the theory defined by the lagrangian \eqref{eq:action8} on
manifolds of $\Spin(7)$ holonomy.

\subsection{Super Yang--Mills on manifolds of $\Spin(7)$ holonomy}
\label{sec:Spin7}

We will fix once and for all a conjugacy class of $\Spin(7)$ subgroups
of $\SO(8)$.  We choose the one for which the positive chirality
spinor representation $\Delta_+$ of $\Spin(8)$ decomposes as
$\bigwedge^1 \oplus \bigwedge^0$, where $\bigwedge^1$ is the vector
representation of $\Spin(7)$ and $\bigwedge^0$ is the trivial
1-dimensional representation.  For this $\Spin(7)$ subgroup, the
vector representation of $\SO(8)$ and the negative chirality spinor
representation $\Delta_-$ remain irreducible and isomorphic to the
spinor representation $\Delta$.  We want to write down the theory in
terms of fields transforming according to irreducible representations
of $\Spin(7)$.  For this we will need to introduce some explicit
projectors onto the irreducible representations of $\Spin(7)$ which
occur in the theory.  Moreover since the fields in the theory all come
from irreducible representations of $\Spin(8)$, it will prove
convenient to define these projectors directly in terms of
$\Gamma$-matrices.

Let us introduce a {\em commuting\/} spinor $\theta \in \Delta_+$
invariant under $\Spin(7)$ and normalised to $\theta^t \theta =
1$. Then $\theta\,\theta^t$ is the projector onto the space of
$\Spin(7)$ invariants in $\Delta_+$.  The Fierz rearrangement formula
yields:
\begin{equation*}
\theta\,\theta^t = \tfrac{1}{16} \left( \1 + \Gamma_9 \right) +
\tfrac{1}{16} \tfrac{1}{4!} \theta^t \Gamma^{ijkl} \theta\,
\Gamma_{ijkl}~.
\end{equation*}
The only supersymmetry transformation which will remain a symmetry of
the action on a spin manifold $M$ of $\Spin(7)$ holonomy is the one
which has the parallel spinor as a parameter.  In other words we must
set $\varepsilon_1 = 0$ and $\varepsilon_2 = \varepsilon \theta$,
where $\varepsilon$ is the anticommuting parameter of the symmetry.
Setting $\varepsilon_1 = 0$ already means that $\delta_\varepsilon
\phi_+ = 0$.  The variation of the gauge field is given by
$\delta_\varepsilon A_i = i \varepsilon_2^t \Gamma_i \psi_1 = i
\varepsilon \theta^t \Gamma_i \psi_1$.  Defining
\begin{equation*}
\psi_i \equiv \theta^t \Gamma_i \psi_1~,
\end{equation*}
we can write it as $\delta_\varepsilon A_i = i \varepsilon \psi_i$.
Notice that $\psi_i$ can be interpreted now as a 1-form in $M$.  Its
variation can be readily calculated to give $\delta_\varepsilon \psi_i
= \theta^t \Gamma_i \delta_\varepsilon \psi_1 = D_i\phi_+
\varepsilon$, where we have used that $\theta^t \Gamma_{ij} \theta =
0$.

Under $\Spin(7)$ the spinor $\psi_2 \in \Delta_+$ decomposes into a
singlet and a vector.  As fields on $M$, the singlet $\eta$ is a
fermionic scalar whereas the vector can be understood as a fermionic
2-form $\chi_{ij}$ satisfying an `anti-self-duality' condition to be
defined presently.  We first turn to the scalar.  Clearly, $\eta =
\theta^t \psi_2 = \psi_2^t \theta$, from where it follows that
$\delta_\varepsilon \phi_- = -2 i \varepsilon \eta$.  In turn its
variation can be readily calculated: $\delta_\varepsilon \eta =
\theta^t \delta_\varepsilon \psi_2 = - \half
[\phi_+,\phi_-]\varepsilon$.  Finally we arrive at the vector
component of $\psi_2$.  Consider $\Gamma_{ij}\theta$.  Since $\theta$
is a $\Spin(7)$-singlet, $\Gamma_{ij}\theta$ transforms under
$\bigwedge^2 \oplus \bigwedge^1$ of $\Spin(7)$.  Since $\psi_2$
transforms according to $\bigwedge^0 \oplus \bigwedge^1$, their inner
product will pick out only the $\bigwedge^1$ component: $\chi_{ij} =
\half \theta^t \Gamma_{ij} \psi_2 =  - \half \psi_2^t \Gamma_{ij}
\theta$.  Not all the components of $\chi_{ij}$ are independent.
Indeed, $\chi_{ij}$ satisfies the following `anti-self-duality'
condition:
\begin{equation}\label{eq:chiasd}
\chi_{ij} = - \tfrac16 \Omega_{ijkl} \chi_{kl}~,
\end{equation}
where $\Omega_{ijkl}$ is a self-dual $\Spin(7)$-invariant 4-form on
$M$ defined by:
\begin{equation*}
\Omega_{ijkl} \equiv  \theta^t \Gamma_{ijkl} \theta~.
\end{equation*}
It follows from this expression that
\begin{equation}\label{eq:Omegasquared}
\Omega_{ijmn} \Omega_{mnkl} = 6 \left( \delta_{ik}\delta_{jl} -
\delta_{il}\delta_{jk} \right) - 4 \Omega_{ijkl}~.
\end{equation}
This allows us to define complementary projectors to decompose the
adjoint representation $\bigwedge^2$ of $\Spin(8)$ into irreducible
representations $\bigwedge^1 \oplus \bigwedge^2$ of $\Spin(7)$.
Indeed let,
\begin{align}
\sP_{ij\,kl} &= -\tfrac18 \theta^t \Gamma_{ij}
\Gamma_{kl}\theta~.\label{eq:p7}\\
&= \tfrac18 \left(\delta_{ik} \delta_{jl} - \delta_{il}
\delta_{jk} - \Omega_{ijkl} \right)\notag
\end{align}
$\sP$ is the projector onto $\bigwedge^1$, whereas $\1 - \sP$ is the
complementary projector onto $\bigwedge^2$.  We can verify that the
fact that $\chi_{ij}$ belongs to $\bigwedge^1$, equivalently $\sP \chi
= \chi$, implies \eqref{eq:chiasd}.  It follows from these expressions
that a 2-form $F_{ij}$ in $\bigwedge^2$ of $\Spin(8)$ belongs to the
vector representation $\bigwedge^1$ of $\Spin(7)$ if and only if
\begin{equation*}
\half \Omega_{ijkl} F_{kl} = - 3 F_{ij}~;
\end{equation*}
whereas it belongs to the adjoint representation $\bigwedge^2$ of
$\Spin(7)$ if and only if
\begin{equation*}
\half \Omega_{ijkl} F_{kl} = F_{ij}~.
\end{equation*}
For $F_{ij}$ the Yang--Mills curvature, these equations are the
instanton equations in eight dimensions studied for the first time in
\cite{CDFN,Ward}.

The variation of $\chi_{ij}$ can now be computed as follows:
$\delta_\varepsilon \chi_{ij}  = \half \theta^t \Gamma_{ij}
\delta_\varepsilon \psi_2 = \tfrac14 F_{kl} \theta^t
\Gamma_{ij}\Gamma_{kl}\theta\varepsilon$.  From \eqref{eq:p7} one sees
that $\delta_\varepsilon \chi_{ij} = -2 (\sP F)_{ij} \varepsilon$ or
explicitly as
\begin{equation*}
\delta_\varepsilon \chi_{ij} = - \half \left( F_{ij} - \half
\Omega_{ijkl} F_{kl} \right)\varepsilon~.
\end{equation*}

Finally we can now rewrite the action in terms of the new fields
$A_i$, $\psi_i$, $\phi_\pm$, $\chi_{ij}$ and $\eta$.  First notice
that we can invert the definitions of the spinor fields:
$\psi_1 = - \psi_i \Gamma_i \theta$ and $\psi_2 = \theta\eta -
\tfrac14 \chi_{ij} \Gamma_{ij} \theta$.  Plugging these expressions
into \eqref{eq:action8} we find
\begin{multline}\label{eq:dw8}
\eL = \tfrac14 \|F_{ij}\|^2 + \half (D_i\phi_+,D_i\phi_-) - \tfrac18
\|[\phi_+,\phi_-]\|^2 - \tfrac{i}{2} (\psi_i,
[\phi_-,\psi_i])\\
{} + 2 i (\chi_{ij},D_i\psi_j) + \tfrac{i}{4} (\chi_{ij},
[\phi_+,\chi_{ij}]) + i (\eta, D_i\psi_i) + \tfrac{i}{2} (\eta,
[\phi_+,\eta])~,
\end{multline}
where $D_i$ now includes not just the gauge field but also the
reduction to $\Spin(7)$ of the spin connection.  This action is
invariant under the following fermionic transformation
\begin{xalignat}{2}\label{eq:Spin7preBRST}
\delta A_i &= i \psi_i & \quad \delta \psi_i &= D_i\phi_+\notag\\
\delta \phi_+ &= 0 & \quad \delta \phi_- &= -2i \eta\notag\\
\delta \eta &= -\half [\phi_+,\phi_-] & \quad \delta \chi_{ij} & =
-\half \left( F_{ij} - \half \Omega_{ijkl} F_{kl}\right)~,
\end{xalignat}
which up to gauge transformations and field equations obeys $\delta^2
= 0$.  We finish with the observation that this action agrees with the
action $S_1$ given by equation (12) in \cite{AOS} under the dictionary
in Table \ref{tab:aos}.

\begin{table}[h!]
\renewcommand{\arraystretch}{1.1}
\begin{tabular}{|>{$}c<{$}|>{$}c<{$}|}
\hline
\multicolumn{1}{|c|}{(12) in \cite{AOS}} &
\multicolumn{1}{c|}{\eqref{eq:dw8} here}\\
\hline\hline
A_\alpha & -A_i\\
\varphi & \phi_+ \\
\lambda & - \phi_- \\
\eta & \eta\\
\psi_\alpha & -\psi_i\\
\chi_{\alpha\beta} & \chi_{ij}\\
\phi_{\alpha\beta\gamma\delta} & -\Omega_{ijkl}\\
\hline
\end{tabular}
\vspace{8pt}
\caption{Comparison with \cite{AOS}.\label{tab:aos}}
\end{table}

\subsubsection{The BRST transformations}\label{sec:BRSTSpin7}

The fermionic symmetry given by equation \eqref{eq:Spin7preBRST} only
squares to zero on shell and up to gauge transformations.  In fact, we
find the following:
\begin{xalignat*}{2}
\delta^2 A_i &= iD_i\phi_+ & \quad \delta^2 \psi_i &=
i[\phi_+,\psi_i]\\
\delta^2 \phi_+ &= 0 & \quad \delta^2 \phi_- &= i[\phi_+,\phi_-]\\
\delta^2 \eta &= i [\phi_+,\eta] & \quad \delta^2 \chi_{ij} & =
-2i \sP\,(D_i\psi_j - D_j\psi_i)~;
\end{xalignat*}
but the equation of motion which follows from varying the action with
respect to $\chi_{ij}$ is given by
\begin{equation*}
\sP\,(D_i\psi_j - D_j\psi_i) = -\half [\phi_+,\chi_{ij}]~,
\end{equation*}
whence $\delta^2\chi_{ij} = i[\phi_+,\chi_{ij}]$ {\em on shell\/}.

It is possible to modify the fermionic symmetry in order to obtain
something which does square to zero on the nose.  The problem that
$\delta$ squares to zero up to gauge transformations is routine to
solve and not terribly relevant for our present purposes.
Nevertheless its solution will be described below for completeness.
The problem that $\chi_{ij}$ has to be put on shell is more relevant.
It is solved by introducing a Lie algebra valued bosonic auxiliary
field $\Lambda_{ij}$ satisfying the same `anti-self-duality' condition
\eqref{eq:chiasd} as $\chi_{ij}$ and redefining:
\begin{equation*}
\delta \chi_{ij} = \Lambda_{ij} \qquad\text{and}\qquad
\delta \Lambda_{ij} = i[\phi_+,\chi_{ij}]~.
\end{equation*}
Now $\delta$ is manifestly nilpotent {\em off shell\/} but only up to
a gauge transformation with parameter $i\phi_+$.  We must now write
the action in terms of the auxiliary field.  We consider first the
following `gauge fermion'
\begin{equation*}
\Psi = -\tfrac14 (\chi_{ij},\Lambda_{ij} + 4 (\sP F)_{ij})~.
\end{equation*}
It is clearly gauge invariant, whence $\delta^2\Psi = 0$.  In other
words,
\begin{equation}\label{eq:qPsi}
\delta\Psi = 2i (\chi_{ij} D_i\psi_j) + \tfrac{1}{4}
(\chi_{ij}, [\phi_+,\chi_{ij}]) - \tfrac14 (\Lambda_{ij}, \Lambda_{ij}
+ 4 (\sP F)_{ij})~,
\end{equation}
is invariant under $\delta$.  Solving for $\Lambda_{ij}$ we see that
$\Lambda_{ij} = -2(\sP F)_{ij}$ and plugging it back into the last
term in \eqref{eq:qPsi} we obtain:
\begin{equation*}
\|(\sP F)_{ij}\|^2 = \tfrac14 \|F_{ij}\|^2 - \tfrac18
\Omega_{ijkl}(F_{ij},F_{kl})~.
\end{equation*}
Calculating a little further we see that the lagrangian \eqref{eq:dw8}
can be rewritten as follows:
\begin{equation*}
\eL = \tfrac18 \Omega_{ijkl}(F_{ij},F_{kl}) + \delta \left( \Psi -
\half (\phi_-, D_i\psi_i + \half [\phi_+,\eta]) \right)~,
\end{equation*}
which shows that $\eL$ is obtained up to a BRST exact term from a
`topological' term as befits a cohomological theory.

This result agrees with \cite{BKS} modulo the fact that their action
is missing some of these terms while at the same time containing some
more fields necessary to fix the gauge.  These are the fields which
solve the other problem with $\delta$: that it squares to zero only up
to gauge transformations.  In order to solve this problem we introduce
a ghost (a Lie algebra valued fermionic field) $c$.  We now define a
new transformation $\delta'$, defined on all fields but $c$ as a gauge
transformation with parameter $c$:
\begin{xalignat}{2}\label{eq:Spin7BRST}
\delta' A_i &= D_i c & \quad \delta' \psi_i &= [c,\psi_i]\notag\\
\delta' \phi_+ &= [c,\phi_+] & \quad \delta' \phi_- &=
[c,\phi_-]\notag\\
\delta' \eta &= [c,\eta] & \quad \delta' \chi_{ij} & =
[c,\chi_{ij}]~,
\end{xalignat}
and then defined on $c$ in such a way that ${\delta'}^2 = 0$: $\delta'
c = \half [c,c]$.  On the other hand, $\delta\delta' + \delta'\delta$
is a gauge transformation with parameter $\delta c$.  Therefore if we
define $\delta c = - i \phi_+$, then the combination $\dd \equiv
\delta + \delta'$ squares to zero off shell on all fields, including
$c$.  In order to give dynamics to the ghost it is necessary to also
introduce an antighost $b$ and in order for $\dd^2 b = 0$ we need to
introduce a so-called Nakanishi--Lautrup auxiliary field.  This field
also serves the dual purpose of fixing the gauge so that the
propagator of the gauge field is well-defined and thus allowing in
principle for the perturbative treatment of the theory.  If we follow
this procedure for the gauge fixing $\nabla_i A_i = 0$ one recovers the
action in \cite{BKS} up to the terms $\tfrac14 \delta (\phi_-,
[\phi_+,\eta])$ which they omit.  These terms are presumably not
important in that the theory gives the same observables with or
without them; but are unavoidable if one reduces from 10-dimensional
super Yang--Mills as advocated here.

\section{Seven dimensions}

In this section we describe the reduction of supersymmetric
Yang--Mills theory from $\MM^{9{+}1}$ to $\EE^7$ and to
7-dimensional manifolds of $G_2$ holonomy.

\subsection{Properties of spinors}

Unlike the 8-dimensional case, there is no isomorphism relating
$\Cl(9,1)$ and $\Cl(7,0)$.  However there exists an isomorphism
$\Cl(7,0) \cong \Cl(8,0)^{\mathrm{even}}$, where this last algebra
refers to the subalgebra of $\Cl(8,0)$ generated by even products of
$\Gamma$-matrices.  This isomorphism induces an embedding $\Spin(7)
\subset \Spin(8)$ under which the half-spin representations
$\Delta_\pm$ of $\Spin(8)$ remain irreducible and isomorphic to the
half-spin representation $\Delta$ of $\Spin(7)$, whereas the vector
representation $\bigwedge^1$ of $\Spin(8)$ decomposes into a vector
and a scalar $\bigwedge^0 \oplus \bigwedge^1$.  We will identify
$\Delta$ with $\Delta_+$ once and for all.  In other words, we think
of $\Cl(7,0)$ {\em as\/} $\Cl(8,0)^{\mathrm{even}}$ and of $\Delta$
{\em as\/} $\Delta_+$.

Explicitly, the isomorphism $\Cl(7,0) \cong \Cl(8,0)^{\mathrm{even}}$
is given as follows: let $i$ run from 1 to 7 and let $\Gamma_i$ and
$\Gamma_8$ denote the $\Gamma$-matrices in eight dimensions.  Define
$\Tilde\Gamma_i = \Gamma_i\Gamma_8$.  A moment's reflection shows that
they generate $\Cl(8,0)^{\mathrm{even}}$, whereas it is evident that
they are $\Gamma$-matrices for $\Cl(7,0)$.  This defines an embedding
$\Spin(7) \subset \Spin(8)$, given infinitesimally by: $\half
\Tilde\Gamma_{ij} = \half \Gamma_{ij}$ but where $i,j$ only run from 1
to 7.  The $\Spin(7)$-isomorphism $\Delta_-\cong \Delta \equiv
\Delta_+$ is then given by $\Gamma_8$.

This course of action allows us to use the results of Section
\ref{sec:8spinors} concerning the reduction from $\Cl(9,1)$ to
$\Cl(8,0)$, and sets the stage for a further reduction to
$\Cl(7,0)$, to which we now turn.

\subsection{Dimensional reduction}\label{sec:7dimensions}

We now dimensionally reduce from $\MM^{9{+}1}$ to $\EE^7$ by dropping
the dependence on $x^8$, $x^9$ and $x^0$; that is $\d_8 = \d_9 = \d_0
= 0$.  This choice of privileged coordinates breaks the Lorentz
symmetry down to $\SO(7) \times \SO(2,1)$.  This suggests that we
arrange the fields into irreducible representations of this group or
its spin cover.  The gauge field will break up into an $\SO(2,1)$
singlet and $\SO(7)$ vector $A_i$ and an $\SO(7)$ singlet and
$\SO(2,1)$ triplet $\phi_\alpha$ where $(\phi_0,\phi_1,\phi_2) =
(A_0,A_9,A_8)$.  The bosonic part of the action defined by
\eqref{eq:9+1SYM} can therefore be written as
\begin{equation*}
\eL_B = \tfrac14 \|F_{ij}\|^2 + \half \eta^{\alpha\beta}
(D_i\phi_\alpha, D_i\phi_\beta) + \tfrac14 \eta^{\alpha\gamma}
\eta^{\beta\delta}([\phi_\alpha,\phi_\beta],[\phi_\gamma,\phi_\delta])~,
\end{equation*}
where $\eta=(-++)$ is the $SO(2,1)$ invariant metric.

As for the fermions, $\Psi = (\lambda_1~\lambda_2)^t$ where
$\lambda_1\in\Delta_-$ and $\lambda_2\in\Delta_+$ of $\Spin(8)$.  We
want to decompose them into representations of $\Spin(7)\times\SL(2)$,
where $\SL(2)$ is the spin cover of $\SO(2,1)$.  Let $\bigodot^p$
denote the $p$-th symmetric tensor product of the defining (real
2-dimensional) representation of $\SL(2)$.  For example,
$\bigodot^0$ is the singlet, $\bigodot^1$ is the defining
representation and $\bigodot^2$ is the adjoint representation.

Let us therefore introduce a doublet of spinors $\psi_A \in
\Delta\otimes\bigodot^1$, defined by: $\psi_1 = \Gamma_8\lambda_1$ and
$\psi_2 = \lambda_2$.  In terms of these fields, the fermionic action
becomes
\begin{equation}\label{eq:LF7}
\eL_F = \tfrac{i}{2} \epsilon^{AB} (\psi_A^t, \Tilde\Gamma_i D_i
\psi_B) + \tfrac{i}{2} \eta^{\alpha\beta} M_\beta^{AB} (\phi_\alpha,
[\psi_A^t,\psi_B])~,
\end{equation}
where the matrices $M_\alpha^{AB}$ are such that
\begin{equation*}
\phi^{AB} \equiv \eta^{\alpha\beta} \phi_\alpha M_\beta^{AB} =
\begin{pmatrix}
\phi_1 - \phi_0 & \phi_2\\
\phi_2 & -(\phi_1+\phi_0)
\end{pmatrix}~.
\end{equation*}
Our conventions for $\epsilon_{AB}$ are as follows.  We take
$\epsilon^{12} = -\epsilon^{21} = \epsilon_{12} = - \epsilon_{21} =
+1$.  We raise and lower indices using the ``northeast'' convention:
\begin{equation*}
O_A = \epsilon_{AB} O^B \qquad\text{and}\qquad
O^A = O_B \epsilon^{BA} = - \epsilon^{AB} O_B~.
\end{equation*}
Therefore ${\epsilon_A}^B = \epsilon_{AC} \epsilon^{CB} = -
\delta_A^B$.

The action inherits the supersymmetry from ten dimensions and in
particular from \eqref{eq:9+1SUSY}.  We let $\varepsilon =
(\xi_1~\xi_2)^t$ where $\xi_1\in\Delta_-$ and $\xi_2\in\Delta_+$.  As
before we introduce a doublet of spinors $\varepsilon_A \in
\Delta\otimes\bigodot^1$ by
\begin{equation*}
\varepsilon_1 = \Gamma_8\xi_1\qquad\text{and}\qquad \varepsilon_2 =
\xi_2~.
\end{equation*}
Decomposing \eqref{eq:9+1SUSY} in terms of irreducible representations
of $\Spin(7) \times \SL(2)$ we find:
\begin{align}\label{eq:7SUSY}
\delta_\varepsilon A_i &= i \epsilon^{AB}\varepsilon_A^t
\Tilde\Gamma_i \psi_B\notag\\
\delta_\varepsilon \phi_\alpha &= i M_\alpha^{AB} \varepsilon_A^t
\psi_B\notag\\
\delta_\varepsilon \psi_A &= \half F_{ij} \Tilde\Gamma_{ij}
\varepsilon_A - D_i{\phi_A}^B\Tilde\Gamma_i \varepsilon_B +
\half {[\phi,\phi]_A}^B \varepsilon_B~,
\end{align}
where
\begin{equation*}
\half [\phi,\phi]^{AB} \equiv \half \epsilon_{CD} [\phi^{AC},\phi^{DB}] =
\begin{pmatrix}
{}[\phi_2,\phi_1 - \phi_0] & [\phi_0,\phi_1]\\
{}[\phi_0,\phi_1] & [\phi_2, \phi_1 + \phi_0]
\end{pmatrix}~.
\end{equation*}
The action is then the sum of $\eL_F$ and the bosonic action $\eL_B$
which can be rewritten in terms of $\phi^{AB}$ as
\begin{equation}\label{eq:LBG2}
\eL_B = \tfrac14 \|F_{ij}\|^2 + \tfrac14 \Tr (D_i\phi,D_i\phi) +
\half \Det\,\half [\phi,\phi]~,
\end{equation}
where $\Tr (D_i\phi,D_i\phi) = (D_i{\phi_A}^B, D_i{\phi_B}^A)$, and
the last term is the `determinant' of the matrix $\half
{[\phi,\phi]_A}^B$ defined above:
\begin{equation*}
\Det\,\half [\phi,\phi] = (\half {[\phi,\phi]_1}^1, \half
{[\phi,\phi]_2}^2) - (\half {[\phi,\phi]_1}^2, \half
{[\phi,\phi]_2}^1)~.
\end{equation*}

\subsection{Reduction of the holonomy group}\label{sec:G2}

In order to define the action $\eL = \eL_B + \eL_F$ on a
7-dimensional spin manifold $M$ instead of $\EE^7$ we need only
covariantise the derivatives to include the spin connection.  As in
eight dimensions, the supersymmetry transformations in
\eqref{eq:7SUSY} will not be a symmetry of the action unless the
parameters are parallel spinors, which reduces the holonomy of $M$
from $SO(7)$ to (a subgroup of) $G_2$.  It therefore pays to rewrite
the theory in terms of irreducible representations of $G_2 \subset
SO(7)$.  Under $G_2$ the spinor representation breaks up as $\Delta
\cong \varrho \oplus \tau$ where $\tau$ is a singlet of $G_2$ and
$\varrho$ is in the 7-dimensional irreducible representation of
$G_2$.  On the other hand, the vector representation $\bigwedge^1$ of
$SO(7)$ remains irreducible and goes over to $\varrho$.

Let $\theta$ denote a commuting parallel spinor in $M$ normalised
pointwise to $\theta^t\theta = 1$.  The parallel spinor will allow us
to decompose $\psi_A$ into its $G_2$ irreducible components.  In other
words, we introduce fields $\chi_{iA}$ and $\eta_A$ which transform
under $G_2\times \SL(2)$ as $\varrho\otimes\bigodot^1$ and
$\tau\otimes\bigodot^1$ respectively.  These fields are defined as
follows: $\chi_{iA} = \theta^t \Tilde\Gamma_i \psi_A$ and $\eta_A =
\theta^t \psi_A$.  Using the fact that $\theta^t \Tilde\Gamma_{ij}
\theta = 0$, we can invert the above definitions and write $\psi_A$ as
follows:
\begin{equation*}
\psi_A = \eta_A \theta - \chi_{iA}\Tilde\Gamma_i\theta~.
\end{equation*}
In terms of these new fields, the fermionic action $\eL_F$ in
\eqref{eq:LF7} is given by:
\begin{multline}\label{eq:LFG2}
\eL_F = i \epsilon^{AB} (\eta_A, D_i\chi_{iB}) - \tfrac{i}{2}
\epsilon^{AB} \varphi_{ijk} (\chi_{iA},D_j\chi_{kB})\\
+ \tfrac{i}{2} (\phi^{AB}, [\eta_A, \eta_B]) + \tfrac{i}{2}
(\phi^{AB}, [\chi_{iA}, \chi_{iB}])~,
\end{multline}
where we have used that $\theta^t\Tilde\Gamma_i\theta =
\theta^t\Tilde\Gamma_{ij}\theta = 0$ and where we have defined
\begin{equation*}
\varphi_{ijk} \equiv \theta^t \Tilde\Gamma_{ijk} \theta~.
\end{equation*}

The action is invariant under two fermionic symmetries, obtained from
\eqref{eq:7SUSY} by choosing the spinors $\varepsilon_A$ to be
parallel.  In other words we let $\varepsilon_A = \lambda_A \theta$
where $\lambda_A$ are anticommuting parameters.  Rewriting
\eqref{eq:7SUSY} in terms of the new fields and for this choice of
supersymmetry parameters, we find:
\begin{xalignat}{2}\label{eq:BRSTG2}
\delta_\lambda A_i &= i \epsilon^{AB} \lambda_A \chi_{iB} & \quad
\delta_\lambda \phi_\alpha & = i M_\alpha^{AB}\lambda_A\eta_B\notag\\
\delta_\lambda \eta_A & = \half {[\phi,\phi]_A}^B \lambda_B & \quad
\delta_\lambda \chi_{iA} & = \half \varphi_{ijk} F_{jk} \lambda_A +
D_i{\phi_A}^B \lambda_B~,
\end{xalignat}
where we have again used that $\theta^t\Tilde\Gamma_i\theta =
\theta^t\Tilde\Gamma_{ij}\theta = 0$.

The 3-form $\varphi_{ijk}$ defined above is a $G_2$ singlet in
$\bigwedge^3\varrho$.  We can relate this to the $\Spin(7)$-invariant
4-form $\Omega_{ijkl}$ defined in Section \ref{sec:Spin7}.  Under the
identification $\Delta \equiv \Delta_+$, the normalised
$G_2$-invariant spinor $\theta$ is (up to a sign) the
$\Spin(7)$-invariant spinor introduced in Section \ref{sec:Spin7} and
bearing the same name.  Therefore,
\begin{equation*}
\varphi_{ijk} = \theta^t \Tilde\Gamma_{ijk} \theta = \theta^t
\Gamma_{ijk8} \theta = \Omega_{ijk8}~.
\end{equation*}
This 3-form obeys an identity similar to the relation
\eqref{eq:Omegasquared} obeyed by $\Omega_{ijkl}$ and derivable in the
same way:
\begin{equation}\label{eq:G2identity}
\varphi_{ijm}\varphi_{mkl} = \delta_{ik}\delta_{jl} -
\delta_{il}\delta_{jk} - \Tilde\varphi_{ijkl}~,
\end{equation}
where the $G_2$-invariant 4-form $\Tilde\varphi$ is simply the
restriction of $\Omega$:
\begin{equation*}
\Tilde\varphi_{ijkl} = \theta^t \Tilde\Gamma_{ijkl} \theta = \theta^t
\Gamma_{ijkl} \theta = \Omega_{ijkl}~.
\end{equation*}
One can check that $\Tilde\varphi$ is the 7-dimensional
Hodge dual of $\varphi$.  Two more identities relate $\Tilde\varphi$
and $\varphi$:
\begin{align}\label{eq:G2identities}
\Tilde\varphi_{ijmn} \Tilde\varphi_{mnkl} &= 4 \left(
\delta_{ik}\delta_{jl} - \delta_{il}\delta_{jk} \right) - 2
\Tilde\varphi_{ijkl}\notag\\
\Tilde\varphi_{ijmn}\varphi_{mnk} &= -4 \varphi_{ijk}~.
\end{align}
These identities are consistent with \eqref{eq:Omegasquared} and
indeed, together with \eqref{eq:G2identity}, imply it.

Just as in the case of $\Spin(7)$ holonomy it is possible to use the 4-form
$\Tilde\varphi$ in order to construct projectors onto the $G_2$ irreducible
subspaces of the adjoint representation $\bigwedge^2$ of $\SO(7)$.  Under
$G_2$, this representation breaks up as $\varrho \oplus \fg_2$, where
$\fg_2$ denotes the (fourteen-dimensional) adjoint representation of $G_2$.
The 3-form $\varphi$ defines a $G_2$-equivariant map $\bigwedge^1 \to
\bigwedge^2$ by $v_i \mapsto \varphi_{ijk} v_k$ which sets up an
isomorphism onto the subrepresentation $\varrho \subset \bigwedge^2$.
It also allows us to define another $G_2$-equivariant map $\bigwedge^2
\to \bigwedge^1$ by $u_{ij} \mapsto \half \varphi_{ijk} u_{jk}$ whose
kernel coincides with the subrepresentation $\fg_2 \subset
\bigwedge^2$.  Using these facts and the identities
\eqref{eq:G2identity} and \eqref{eq:G2identities} the expressions for
the projectors follow easily.  For the projector $\sP$ onto $\varrho$
we find
\begin{equation*}
\sP_{ij\,kl} = \tfrac16 \left(\delta_{ik} \delta_{jl} - \delta_{il}
\delta_{jk} - \Tilde\varphi_{ijkl} \right)~,
\end{equation*}
and naturally $\1 - \sP$ for the projector onto $\fg_2$.  It follows
from these expressions that a 2-form $F_{ij} \in \bigwedge^2$ belongs
to the subrepresentation $\varrho$ if and only if
\begin{equation*}
\half \Tilde\varphi_{ijkl} F_{kl} = - 2 F_{ij}~;
\end{equation*}
whereas it belongs to the subrepresentation $\fg_2$ if and only if
\begin{equation*}
\half \Tilde\varphi_{ijkl} F_{kl} = F_{ij}~.
\end{equation*}
As in eight dimensions, for $F_{ij}$ the Yang--Mills curvature, these
equations are the instanton equations in seven dimensions defined for
the first time in \cite{CDFN,Ward}.

\subsection{A cohomological theory for instantons}
\label{sec:G2instanton}

The theory described in Section \ref{sec:G2} has two BRST symmetries:
one for each parameter $\lambda_A$.  In order to fix a BRST operator,
we set $\lambda_1=0$ and let $\lambda_2 = -\lambda$.  Any other choice
will be related to this one by an $\SO(1,1)\subset\SL(2)$
transformation, but this is an automorphism of the resulting theory.
We let $\phi \equiv \phi_2$ and $\phi_\pm = \phi_1 \pm \phi_0$.  We
further define $\psi_i = \chi_{i1}$ and $\chi_i = \chi_{i2}$.  In
terms of these fields the fermionic symmetry in \eqref{eq:BRSTG2}
(omitting the parameter $\lambda$) becomes:
\begin{xalignat}{2}\label{eq:G2BRST}
\delta A_i &= i\psi_i & \quad
\delta \psi_i & =  D_i \phi_+ \notag\\
\delta \phi & = -i\eta_1 & \quad
\delta \eta_1 & = [\phi,\phi_+]\notag\\
\delta \phi_- & = 2i \eta_2 & \quad
\delta \eta_2 & = \half [\phi_+,\phi_-] \notag\\
\delta \phi_+ & = 0 & \quad
\delta \chi_i & = D_i\phi - \half \varphi_{ijk} F_{jk}~.
\end{xalignat}

The action is given by $\eL_B + \eL_F$, which are in turn given by
equations \eqref{eq:LBG2} and \eqref{eq:LFG2}.

As before $\delta^2$ only squares to zero on the $\chi_i$ shell and
modulo a gauge transformation with parameter $i\phi_+$.  We can remedy
this situation as we did in Section \ref{sec:BRSTSpin7}.  In order to
lift the on-shell condition we introduce a bosonic Lie algebra valued
auxiliary field $\Lambda_i$ and define $\delta \chi_i = \Lambda_i$ and
$\delta \Lambda_i = i[\phi_+,\chi_i]$.  It is now clear that $\delta$
squares off-shell to a gauge transformation with parameter $i\phi_+$.
Introducing as before a ghost $c$ with $\delta c = -i\phi_+$ and a
second fermionic symmetry $\delta'$ defined as a gauge transformation
with parameter $c$ on all fields but $c$ itself and by $\delta' c =
\half [c,c]$, guarantees that the combined fermionic symmetry $\dd =
\delta + \delta'$ now squares to zero off shell.  At the end of the
day the lagrangian can be written as the gauge fixing of a topological
term:
\begin{equation}\label{eq:LG2}
\eL = \tfrac18 \Tilde\varphi_{ijkl} (F_{ij},F_{kl}) + \half
\varphi_{ijk} (D_i\phi,F_{jk}) + \dd \Psi~,
\end{equation}
where the gauge fermion $\Psi$ is now given by
\begin{equation}\label{eq:GFG2}
\Psi = -\half (\chi_i, \Lambda_i - 2 D_i\phi + \varphi_{ijk}F_{jk}) -
\half (\phi_-, D_i\psi_i + [\phi,\eta_1] - \half [\phi_+,\eta_2])~.
\end{equation}

At first sight the above theory seems to describe the topology of what
could be termed the monopole moduli space on a 7-manifold of $G_2$
holonomy; but as it stands it computes invariants of the moduli space
of instantons.  To see this we discuss first some extra structure that
this theory possesses when we take both BRST symmetries into
consideration.

\subsection{A balanced cohomological field theory}

The theory described in Section \ref{sec:G2} actually has a richer
structure than the one just described.  If instead of focusing on one
of the BRST operators we take both into account we are led to a
structure which the authors of \cite{DM} call a {\em balanced
topological field theory\/}.  As shown in \cite{BTNT2too} balanced
topological field theories are in fact equivalent to a class of
topological theories possessing two topological charges
\cite{BTSQM,BTNT2}.  Balanced topological field theories are
characterised by having two BRST operators and a global $\SL(2)$
symmetry.  It is remarkable that this is precisely the structure
present in the dimensional reduction of 10-dimensional super
Yang--Mills theory on a riemannian 7-manifold of $G_2$ holonomy.  One
may be tempted to think that this example is paradigmatic: giving as
it does a geometric origin for the $\SL(2)$ symmetry.

Let us define an $\SL(2)$ doublet of fermionic transformations
$\delta_A$ by $\delta_\lambda = \epsilon^{AB} \lambda_A \delta_B$,
where $\delta_\lambda$ is defined by equation \eqref{eq:BRSTG2}.  From
this equation we can then read off the action of $\delta_A$ on the
fields:
\begin{xalignat}{2}\label{eq:BTFT}
\delta_A A_i &= i \chi_{iA} & \quad
\delta_A \phi_{BC} & = i (\epsilon_{AB} \eta_C + \epsilon_{AC}
\eta_B)\notag\\
\delta_A \eta_B & = -\half [\phi,\phi]_{AB} & \quad
\delta_A \chi_{iB} & = -\half \varphi_{ijk} F_{jk} \epsilon_{AB} -
D_i\phi_{AB}~.
\end{xalignat}
It is routine to prove that up to equations of motion,
$\{\delta_A,\delta_B\}$ equals a gauge transformation with parameter
$-2i\phi_{AB}$.  As usual, this is off shell for all fields but
$\chi_{iA}$.  In order to lift this restriction we will introduce an
auxiliary field $\Lambda_i$ and redefine
\begin{equation*}
\delta_A \chi_{iB} = \epsilon_{AB} \Lambda_i - D_i\phi_{AB}
\qquad\text{and}\qquad
\delta_A \Lambda_i = -i [{\phi_A}^B,\chi_{iB}] + i D_i\eta_A~.
\end{equation*}
Notice that $\{\delta_A,\delta_B\} = -2i[\phi_{AB},-]$ on both
$\chi_{iA}$ and $\Lambda_i$.  An important feature of balanced
topological field theories is that the action is given in terms of a
gauge invariant potential.  In this case we have the following
expression for the action:
\begin{equation}\label{eq:LBTFTG2}
\eL = \eL_{\mathrm{top}} + \half \epsilon^{AB} \delta_A\delta_B \eV~,
\end{equation}
where
\begin{align}
\eL_{\mathrm{top}} &= \tfrac18 \Tilde\varphi_{ijkl} (F_{ij},F_{kl})\\
\intertext{and}
\eV &= \tfrac{i}{2} \CS(A) - \epsilon^{AB}
(\chi_{iA},\chi_{iB}) + \tfrac12 \epsilon^{AB} (\eta_A,\eta_B) ~,
\end{align}
where $\CS(A)$, reminiscent of the Chern--Simons form, is given by:
\begin{align}\label{CSForm}
\CS(A) &= \varphi_{ijk}\left( (A_i,\d_j A_k) - \tfrac13 (A_i,[A_j,A_k])
\right)\notag\\
&= \half \varphi_{ijk}\left( (A_i,F_{jk}) + \tfrac13 (A_i,[A_j,A_k])
\right)~.
\end{align}
As with the genuine Chern--Simons form, this $\CS(A)$ is invariant
under infinitesimal gauge transformations, which is all that is
required for $\eL$ to be BRST-invariant.  Balanced topological field
theories have the property that the path integral localises on the
critical points of the potential.  In this case, the critical points
of the potential $\eV$ are configurations for which $\chi_{iA} =
\eta_A = 0$ and for which the Yang--Mills curvature defines an
instanton: $\varphi_{ijk} F_{jk} = 0$.  It follows that this theory
computes topological invariants of the instanton moduli space.  Its
partition function, for example, computes the Euler characteristic.

\subsection{Comparing with \cite{AOS}}

Now we compare the action given by \eqref{eq:LBG2} and \eqref{eq:LFG2}
with the one in \cite{AOS}, or equivalently with the action in Section
\ref{sec:Spin7} reduced to seven dimensions and to holonomy $G_2$.  In
order to do this we will have to truncate the theory and at the same
time rewrite it in terms of a slightly different set of fields.  We
first set $\phi_2$ (that is, $A_8$) to zero.  Setting its variation to
zero demands that we take $\lambda_1\eta_2 = - \lambda_2\eta_1$.
Without loss of generality we will take $\lambda_1 = \eta_1 = 0$ and
$\lambda_2 \equiv - \lambda$ and $\eta_2 \equiv -\eta$.  Any other
choice will be related to this one by an $\SO(1,1)\subset\SL(2)$
transformation, but this is an automorphism of the resulting theory.
This leaves only one remaining fermionic symmetry which, omitting the
parameter $\lambda$, is given by:
\begin{xalignat}{2}\label{eq:G2truncatedBRST}
\delta A_i &= i \chi_{i1} & \quad
\delta \eta & = -[\phi_0,\phi_1] \notag\\
\delta \phi_0 & = i\eta & \quad
\delta \chi_{i1} & = D_i (\phi_0 + \phi_1)\notag\\
\delta \phi_1 & = -i\eta & \quad \delta \chi_{i2} & =
-\half \varphi_{ijk} F_{jk}~.
\end{xalignat}
By inspection, we find that under the following dictionary:
\begin{equation}\label{eq:G2dictionary}
\phi_\pm = \phi_1 \pm \phi_0~,\qquad
\psi_i = \chi_{i1}\qquad\text{and}\qquad
\chi_{ij} = \half \varphi_{ijk} \chi_{k2}~,
\end{equation}
and after using identity \eqref{eq:G2identity}, the two sets of
transformation laws \eqref{eq:Spin7preBRST} and
\eqref{eq:G2truncatedBRST} agree.  Moreover, a little calculation
shows that in terms of these fields the action agrees with the
dimensional reduction of \eqref{eq:dw8} and the truncation $A_8 =
\psi_8 = 0$.  The only subtlety in this calculation is the fact that
in \eqref{eq:dw8}, $\chi$ has components $\chi_{ij}$ and $\chi_{i8}$.
But using the `anti-self-duality' condition \eqref{eq:chiasd}, we can
solve for $\chi_{i8}$ in terms of $\chi_{ij}$: $\chi_{i8} = -\tfrac16
\varphi_{ijk}\chi_{jk}$.  In addition $\chi_{ij}$ also obeys its own
`anti-self-duality' condition $\chi_{ij} = -\tfrac14
\Tilde\varphi_{ijkl} \chi_{kl}$, as can be read from the second
identity in \eqref{eq:G2identities} and definition of $\chi_{ij}$ in
\eqref{eq:G2dictionary}.  This is to be expected since by construction
$\chi_{ij}$ belongs to the subrepresentation $\varrho$.  Finally to
make contact with the action in \cite{AOS} it is necessary to further
rescale $\chi_{ij}$: $\chi_{ij}^{\text{\cite{AOS}}} = -\tfrac43
\chi_{ij}$.

It should be mentioned, however, that this truncation results in only
a partial gauge-fixing of the `topological' symmetries of the original
theory, since the field $\eta$ is no longer propagating.  Therefore
the theory is not well-defined.  The correct cohomological theory
describing the topology of the instanton moduli space for 7-manifolds
of $G_2$ holonomy is the untruncated theory given by the sum of
\eqref{eq:LBG2} and \eqref{eq:LFG2}, or equivalently by
\eqref{eq:LBTFTG2}.

\subsection{Monopoles in seven dimensions}

The BRST transformation law for $\chi_i$ in \eqref{eq:G2BRST} is
reminiscent, when set to zero, of an equation of Bogomol'nyi type.
In fact, this equation has certain parallels with the more familiar
case of 3-di\-men\-sion\-al monopoles, which we would like to
exploit. Let us then first review the 3-dimensional case.  A
monopole configuration is specified by a gauge field $A_i$ and an
adjoint Higgs $\phi$ satisfying the Bogomol'nyi equation
\begin{equation}\label{eq:bogeqn3}
D_i \phi = \pm \half \epsilon_{ijk} F_{jk}~,
\end{equation}
where the sign differentiates the monopole from the antimonopole.  We
can understand this equation as a (anti-)self-duality equation in four
dimensions by simply thinking of the Higgs field as the fourth
component of a 4-dimensional gauge field $A_{\underline{i}} =
(A_i,\phi)$.  Of course, the fourth dimension is fake and
$A_{\underline{i}}$ does not depend on it.  Nevertheless the
self-duality equation
\begin{equation*}
F_{\underline{ij}} = \pm \half \epsilon_{\underline{ijkl}} 
F_{\underline{kl}}~, 
\end{equation*}
is clearly seen to be equivalent to the Bogomol'nyi equation
\eqref{eq:bogeqn3}. 

Now let us consider the 7-dimensional case.  In order to guess the
form of the Bogomol'nyi equation, we play the same game and
dimensionally reduce the 8-dimensional instanton equations.  Again we
have a gauge field $A_i$ and a Higgs field $\phi$, but where $i$ runs
from 1 to 7, and we will understand $\phi$ as $A_8$ but with the tacit
understanding that $A_{\underline{i}} = (A_i,\phi)$ does not depend on
$x^8$.  The 8-dimensional curvature has components $F_{ij}$ and
$F_{i8} = D_i\phi$.  As we saw above there are two possible instanton
equations in eight dimensions.  We can demand that $F$ belongs either
to $\bigwedge^1$ or to $\bigwedge^2$ of $\SO(7)$.  It is the latter
equation with which this paper is concerned; although it would be
interesting to investigate the other equations and in particular its
possible supersymmetric origins.\footnote{Notice that the other
8-dimensional instanton equation is equivalent to the equation:
$F_{ij} = - \varphi_{ijk} D_k\phi$; which is in turn equivalent to one
of the instanton equations in seven dimensions. This is to be expected
since both instanton equations restrict the curvature to the
$\bigwedge^1$ of $\Spin(7)$ which remains irreducible under $G_2$.}
Demanding that $F$ belongs to $\bigwedge^2$ is equivalent to
\begin{equation*}
\half\Omega_{\underline{ijkl}} F_{\underline{kl}} =
F_{\underline{ij}}~.
\end{equation*}
Remarkably, as in three dimensions, this equation is {\em
equivalent\/} to the equation of Bogomol'nyi type:
\begin{equation}\label{eq:bogeqn7}
D_i\phi = \half \varphi_{ijk} F_{jk}~.
\end{equation}
Notice parenthetically that this equation reduces to one of the
instanton equations when the 7-manifold is compact: just as in three
dimensions, using the Bianchi identity, the Bogomol'nyi equation
implies the equation $D_iD_i\phi=0$, which in a compact space implies
$D_i\phi=0$.

Now let $X\in\fg$ be a fixed element of the Lie algebra, and consider
the gauge-fixing condition: $\nabla_i A_i = [X,\phi]$.  We can incorporate
this condition on the action in the usual way.  We introduce the
antighost $b$ and the Nakanishi--Lautrup auxiliary field $\Sigma$,
fermionic and bosonic respectively, and both Lie algebra valued.
Their transformation laws are given by:
\begin{equation*}
\delta b = \Sigma \qquad\text{and}\qquad \delta \Sigma = i[\phi_+,b]~,
\end{equation*}
which supplement equation \eqref{eq:G2BRST}.  $\delta'$ on these
fields is defined as gauge transformations with a ghost parameter:
\begin{equation*}
\delta' b = [c,b] \qquad\text{and}\qquad \delta' \Sigma = [c,\Sigma]~.
\end{equation*}
The new action is again given by equation \eqref{eq:LG2}, but where
the gauge fermion $\Psi$ in equation \eqref{eq:GFG2} receives an extra
term: $(b,\Sigma - \nabla_iA_i + [X,\phi])$.  By analogy with the
3-dimensional case, the resulting theory presumably localises on the
solutions of equation \eqref{eq:bogeqn7}.  However, this being a
seven-dimensional theory makes an explicit verification difficult.

\section{Conclusion}

In this paper we have proven the conjectures made in \cite{BKS} and
\cite{AOS} concerning the supersymmetric origin of the cohomological
field theories appearing in those papers.  This gives further evidence
of the fact that the effective worldvolume theories of curved
euclidean D-branes of type {\II} superstring theory are cohomological.
This holds both for D-branes wrapping around calibrated submanifolds
and for those wrapping around the whole manifold (the whole manifold
of course is a trivial case of calibrated submanifold for which the
calibration is the volume form). Furthermore, for the
higher-dimensional theories defined on six-, seven- and
eight-manifolds, the calibrations (the forms, $\Omega$, $\varphi$,
$\Tilde\varphi$ discussed in the present paper, and also the
holomorphic three-form for the Calabi-Yau three-fold) play an integral
role in the definition.  This leads one to expect that these
higher-dimensional theories will be important in the understanding of
the submanifolds and wrapped D-branes, as well as the for the study as
presented here of a D-brane wrapping the entire manifold.

In Tables \ref{tab:dbranes} and \ref{tab:branescan} we present the
euclidean curved D-brane scan for type {\II} superstrings.  The proofs
that most of these resulting theories are cohomological can be found
in the papers \cite{BSV,BTNT2too,BTESYM} and, of course, in this one.

\begin{table}[h!]
\renewcommand{\arraystretch}{1.1}
\begin{tabular}{|>{$}c<{$}|>{$}c<{$}|l|}
\hline
\dim M & \multicolumn{1}{c|}{Holonomy} &
Calibrated submanifolds (dimension)\\
\hline\hline
8 & \Spin(7) & Cayley (4)\\
8 & \SU(4) & complex (2,4,6), special lagrangian (4)\\
8 & \Sp(2) & complex (2,4,6)\\
7 & G_2 & associative (3), coassociative (4)\\
6 & \SU(3) & complex (2,4), special lagrangian (3)\\
4 & \SU(2) & complex (2)\\
\hline
\end{tabular}
\vspace{8pt}
\caption{Calibrated submanifolds of manifolds $M$.\label{tab:dbranes}}
\end{table}

\begin{table}[h!]
\renewcommand{\arraystretch}{1.1}
\begin{tabular}%
{|c||>{$}c<{$}|>{$}c<{$}|>{$}c<{$}|>{$}c<{$}|>{$}c<{$}|>{$}c<{$}|}
\hline
$p{+}1$ & \SU(2) & \SU(3) & \SU(4) & \Sp(2) & G_2 & \Spin(7)\\
\hline\hline
2 & \times &  \times &  \times &  \times &  & \\
3 & & \times &  & & \times &  \\
4 & \times &  \times &  \times\times &  \times &  \times &  \times \\
5 & & & & & & \\
6 & & \times &  \times &  \times &  & \\
7 & & & & & \times &  \\
8 & & & \times &  \times &  & \times \\
\hline
\end{tabular}
\vspace{8pt}
\caption{Euclidean curved D$p$-brane scan organised by holonomy
group.  This table includes the manifold and its calibrated
submanifolds.  There are two kinds of 4-dimensional calibrated
submanifolds in a Calabi--Yau 4-fold: complex and special lagrangian.
\label{tab:branescan}} 
\end{table}

We end by remarking that formula \eqref{eq:LBTFTG2} relates two
topological quantities associated to seven-dimensional gauge theory:
the instanton charge and the Chern--Simons form.  In particular the
appearance of the Chern--Simons form gives further evidence (see
\cite{DT} for another approach) of the existence of a
higher-dimensional Floer theory.  We hope to return to this point in a
future publication.

\begin{ack}
We would like to thank Matthias Blau and George Thompson for their
comments on an earlier version of this paper.  BSA would like to thank
the PPARC for support, and JMF and BS would like to do likewise to the
EPSRC.
\end{ack}

\providecommand{\bysame}{\leavevmode\hbox to3em{\hrulefill}\thinspace}

\end{document}